\def\cm{{\rm\thinspace cm}} 
\def\erg{{\rm\thinspace erg}}

\def\km{{\rm\thinspace km}} 
 
\def\Mpc{{\rm\thinspace Mpc}} 
 
\def\Zsun{\hbox{$\rm\thinspace Z_{\odot}$}} 
 
\def\s{{\rm\thinspace s}} \def\yr{{\rm\thinspace yr}}

\def\cmsqps{\hbox{$\cm^2 \s^{-1}\,$}}

\def\ergps{\hbox{$\erg\s^{-1}\,$}}

\def\kmps{\hbox{$\km\s^{-1}\,$}}

\def\kmpspMpc{\hbox{$\kmps\Mpc^{-1}$}}

\documentclass[usegraphicx]{mn2e}

\usepackage{amssymb}
\usepackage{mathptmx}

\include{defn}

\begin{document}

\voffset=-0.4in

\title[A deep Chandra observation of the Centaurus cluster]
{A deep \emph{Chandra} observation of the Centaurus cluster: bubbles, 
filaments and edges}
\author[A.C. Fabian et al]
{A.C. Fabian$^1$\thanks{E-mail: acf@ast.cam.ac.uk},
  J.S. Sanders$^1$,  G.B Taylor$^{2,3}$ and S.W. Allen$^{1,2}$\\
  $^1$ Institute of Astronomy, Madingley Road, Cambridge CB3 0HA\\
  $^2$ Kavli Institute for Particle Astrophysics and Cosmology,
Stanford University, 382 Via Pueblo Mall, Stanford, CA 94305-4060, USA\\
  $^3$ National Radio Astronomy Observatory, Socorro, NM 87801, USA}

\maketitle

\begin{abstract} X-ray images and gas temperatures taken from a deep
$\sim$200~ks Chandra observation of the Centaurus cluster are
presented. Multiple inner bubbles and outer semicircular edges are
revealed, together with wispy filaments of soft X-ray emitting gas.
The frothy central structure and eastern edge are likely due to the
central radio source blowing bubbles in the intracluster gas. The
semicircular edges to the surface brightness maps 32~kpc to the east
and 17.5~kpc to the west are marked by sharp temperature increases and
abundance drops. The edges could be due to sloshing motions of the
central potential, or are possibly enhanced by earlier radio activity.
The high abundance of the innermost gas (about 2.5 times Solar) limits
the amount of diffusion and mixing taking place. \end{abstract}


\section{Introduction}

The Centaurus cluster (Abell~3526) is X-ray bright, being the nearest
cluster (redshift $z=0.0104$) with a 2--10~keV luminosity exceeding
$5\times 10^{43}\ergps$. Our earlier 31.7~ks Chandra image of the
Centaurus cluster revealed a complex structure in the innermost few
arcmin of the core, centred on the brightest cluster galaxy NGC\,4696
(Sanders \& Fabian 2002). The iron abundance of the gas was found to
peak at a radius of about 1 arcmin from the centre. The temperature
drops from 3.5 to about 1~keV over this whole region. A plume-like
structure swirls clockwise to the NE beyond which there is an abrupt
temperature increase (i.e. a cold front). The central X-ray emission
is surrounded by marked dips in emission, or bubbles, which coincide
with the complex radio source (Taylor, Fabian \& Allen 2002).

\begin{figure}
  \includegraphics[width=\columnwidth]{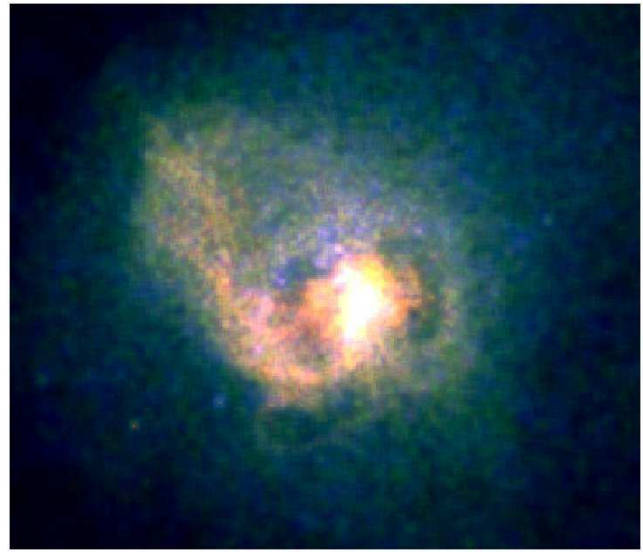}
  \caption{Three colour image of the core of the cluster (24~kpc from
   N to S).  Emission
    between 0.3 and 1~keV is coloured red, 1 to 2~keV green, and 2 to
    7~keV blue. The image is 2~m 21~s from N to S which is almost
30~kpc at the distance of NGC\,4696.}
\end{figure}

\begin{figure*} 
  \includegraphics[width=0.9\textwidth]{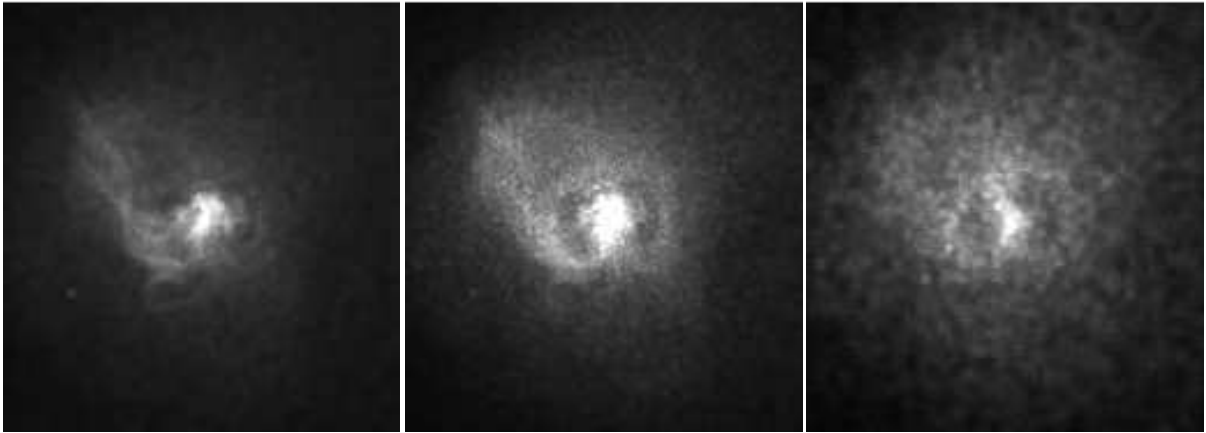}
  \caption{Accumulatively-smoothed images of the core of the cluster
    in three energy bands. 0.3-1.0 (left), 1.0-2.0 (centre) and
    2.0-7.0 (right). The images were smoothed to include a signal to
    noise ratio of 10 in the smoothing kernel. Each image is 2~m 46~s
from N to S.}
\end{figure*}

\begin{figure}
\includegraphics[width=0.95\columnwidth]{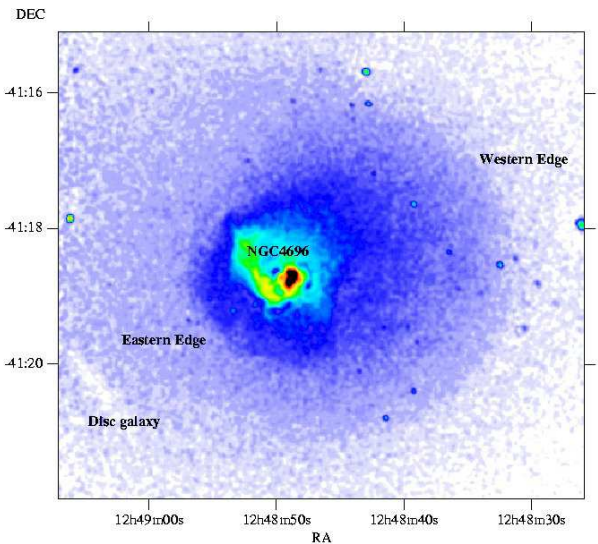}
    \caption{Upper: Detail of the X-ray image in the outer part of the
    core. The image is in the 0.4 to 7~keV band, smoothed with a
    Gaussian of $\sim 1$~arcsec. Using an image smoothed with a
    Gaussian of $\sim 15$~arcsec, we subtracted some of the larger
    scale structure to improve contrast. Various features are marked.
} 
\end{figure}

Previous X-ray observations (e.g. Allen \& Fabian 1994) show a system
with smooth, elliptical, X-ray isophotes, indicating the system is
relatively relaxed. However, there is evidence for a current or past
merger event (Allen \& Fabian 1994; Churazov et al. 1999; Furusho et
al. 2001; Dupke et al 2001) in the form of shifts in X-ray isophote
centroids with radius and bulk motions in the X-ray gas. A
neighbouring subcluster, Cen~45 centred on NGC\,4709 which is about 15
arcmin E of NGC\,4696, has a velocity which is 1500~\kmps higher than
the main Centaurus cluster, Cen~30 (Lucey, Currie \& Dickens 1986).
Observations of the Centaurus cluster using \emph{ROSAT} and
\emph{ASCA} show that the central region of the cluster is
particularly rich in metals, with a large abundance gradient (Fukazawa
et al. 1994; Ikebe et al. 1998; Allen et al. 2001).

Cluster cores are in detail complex but provide us with an observable
analogue of the cooling and heating processes implicit in the
formation of massive galaxies. The nearness, intermediate temperature,
short radiative cooling time and high metallicity make the Centaurus
cluster an excellent candidate for studying these processes and also
the enrichment of the intracluster gas. Here we present images of the
Centaurus cluster from a recent 200~ks Chandra observation.

We adopt $H_0=70\kmpspMpc$ which means that one arcsec corresponds to
210~pc at the redshift of the Centaurus cluster.

\section{The Data}
The data presented here are based on \emph{Chandra} OBSIDs 504, 5310,
4954 and 4955. OBSID 504 was first presented in Sanders \& Fabian
(2002). The standard \textsc{lc\_clean} tool was used to remove
periods in the observations with possible flares yielding a total
good time of 199.3~ks. Each of the datasets were reprocessed to have
the latest gain file, and time dependent gain correction was
applied. We used standard blank sky observations to create background
spectra for use in spectral fitting.

A 3-band X-ray image of the central core is shown in Fig.~1, with the
separate band images in Fig.~2. The images here have been
accumulatively smoothed (Sanders et al in preparation; smoothing with
a circular top hat kernel with radius determined so that the minimum
signal to noise is constant). A whole-band image showing the outer
parts is in Fig.~3 and temperature and abundance maps in Fig.~4. The
distribution of abundances (scaled to those of Anders \& Grevesse
1989) as a function of temperature is shown in Fig.~5. X-ray and
optical images of a disc galaxy seen in X-ray absorption are in
Fig.~6.

An existing VLA 1.4~GHz image has been overlaid on a whole band X-ray
image in Fig~7. The X-ray image has been adaptively-smoothed using an
algorithm due to H. Ebeling; features should be significant above the
3-sigma level.  New VLA observations were obtained on 2004 October 24
with the VLA in its 'A' configuration.  At the observed frequency of
326 MHz this yielded an angular resolution of 17.7 $\times$ 5.6
arcsecond in postion angle $-$4.3~deg.  The total time on source was
178 min.  The bandwidth used was 12.5 MHz in a 4 IF spectral line mode
so as to allow for interference excision using the AIPS task FLGIT.
Calibration and imaging of the data were performed in the standard way
within AIPS.  The final image has an rms noise of 4.8 mJy/beam.  This
image (Fig.~8, Top) compares well with a 1.4 GHz VLA image previously
published (Fig.~7, see also Taylor et al. 2002) and shows that the
radio emission extends nearly 2 arcmin (25 kpc) to the south of
the nucleus.  At the extremities the spectrum is quite steep with a
power law index of $-$1.5 (Fig.~8, Lower).  

Strong Faraday Rotation is observed in the central radio source
indicating a mean magnetic field of about $8\mu{\rm G}$ at 10~kpc
radius (Taylor et al 2002). This corresponds to a magnetic pressure
there which is about 2 per cent of the thermal pressure.

\begin{figure*} \includegraphics[width=\textwidth]{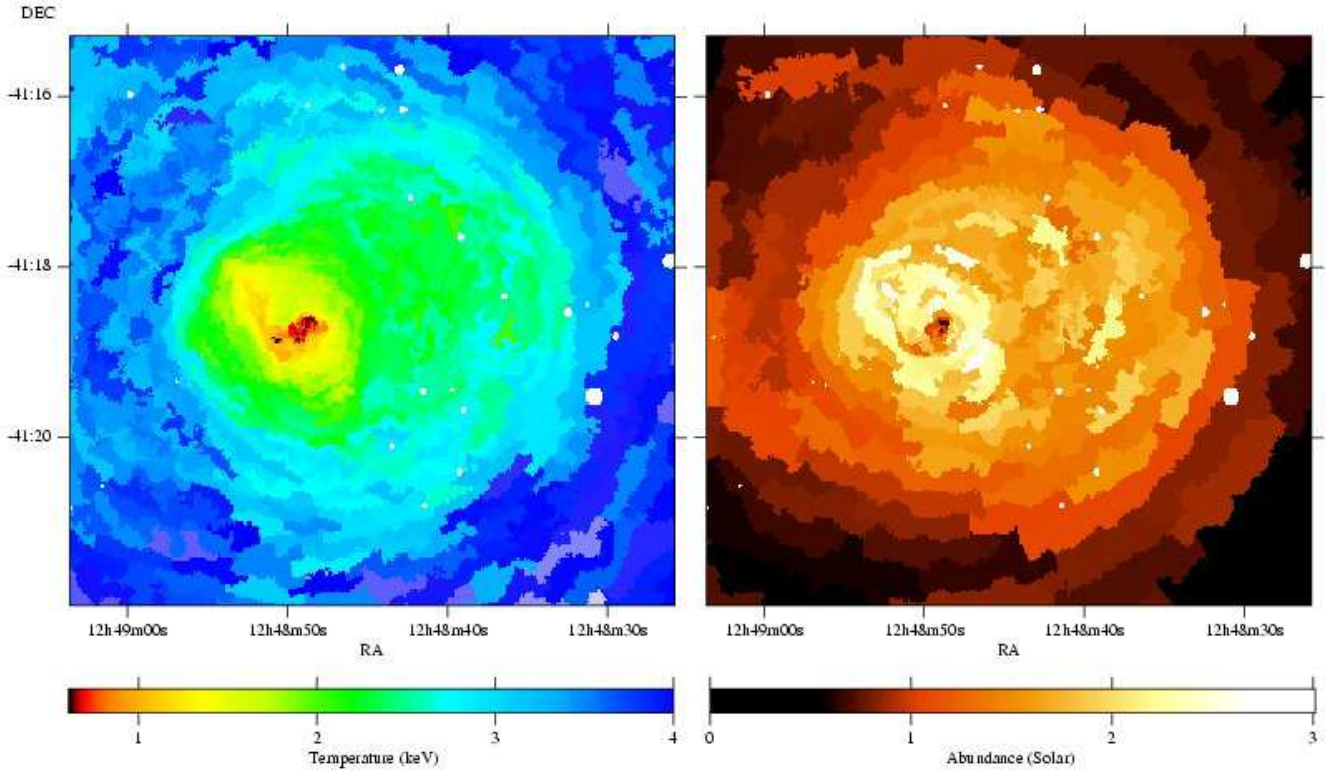} 
    \caption{ Left: Projected, emission-weighted, temperature map of
    the cluster. Excluded point sources are marked by white discs.
    The uncertainties on the temperatures in individual regions
    containing 2500~counts vary from $\sim 0.01$~keV in the coolest
    regions to around $0.5$~keV in the hottest regions. Right:
    Corresponding abundance map measured by fitting spectra in regions
    containing $\gtrsim 10^4$ counts. Two-temperature models were
    tested for each region and, where required by an F-test (at better
    than the 99 per cent level), we have replaced the single
    temperature abundance by the corresponding two-temperature one.
    Two temperatures were need for most of the regions in the
    temperature map below 2~keV. } 
\end{figure*}

\begin{figure}
  \includegraphics[width=0.9\columnwidth]{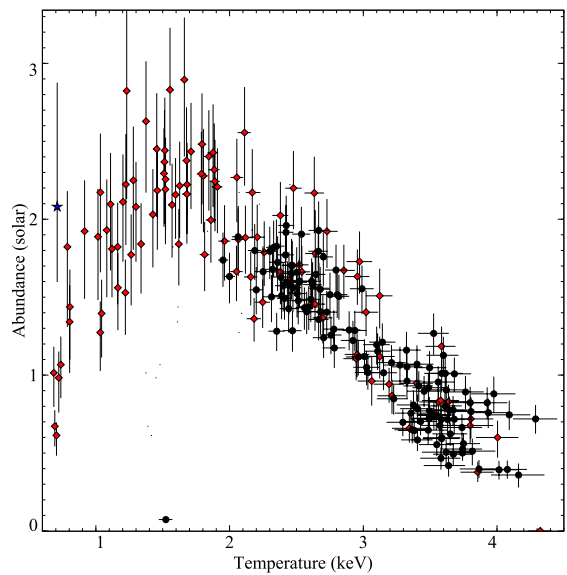}
  \caption{Distribution of abundance as a function of temperature. 
Abundances obtained from a two-temperature fit are plotted in red using 
temperatures from the single-temperature fit. The result of using a
three-temperature model for the apparent low abundance regions below
0.8~keV is shown by the blue star. }
\end{figure}

\section{Results}

The longer X-ray exposure has enabled us to see more deeply into the
cluster core and resolve much more detail. New features include:

$\bullet$ Filaments extending to the E and NE from the centre are
found in the soft X-ray image below 1~keV (Figs.~1 and 2 left). The
inner parts of the filaments correspond to the optical filaments and
dust lane seen in NGC\,4696 (Fabian et al 1982; Sparks, Macchetto \&
Golombek 1989). Comparison with new H$\alpha$ images of this region
will be presented elsewhere (Crawford et al., in preparation). 

$\bullet$ In the 1--2~keV band the holes corresponding to the radio lobes are
very clear and above 2~keV the rims of these `holes' or `bubbles'
appear bright.  The rims do not appear in projection to be hotter
(Fig.~4; confirmed by a higher resolution temperature map) and are
therefore not shocked. This is similar to results on bubbles found in
the Perseus cluster (Fabian et al 2002, 2003). Possible weak ripples
in the X-ray surface brightness centred on the nucleus can be seen in
Fig.~3, particularly on the NE of the E edge (cf. Fabian et al 2003;
Forman et al 2003).


\begin{figure}
  \includegraphics[width=\columnwidth]{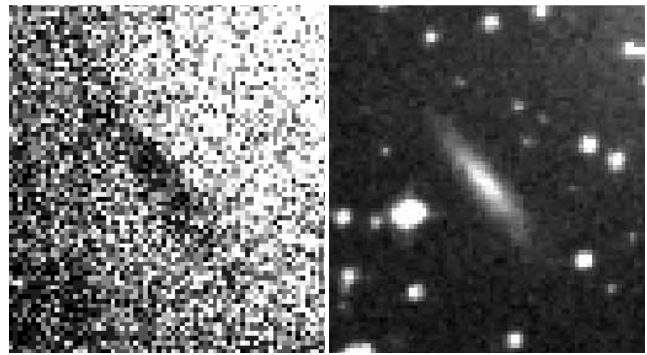}
  \caption{Soft (0.3--0.8~keV) X-ray image (left) of the absorbing disc 
  galaxy seen in the red DSS image (right). Each image is 2.5~arcmin
from N to S. }
\end{figure}

\begin{figure}
  \includegraphics[width=\columnwidth]{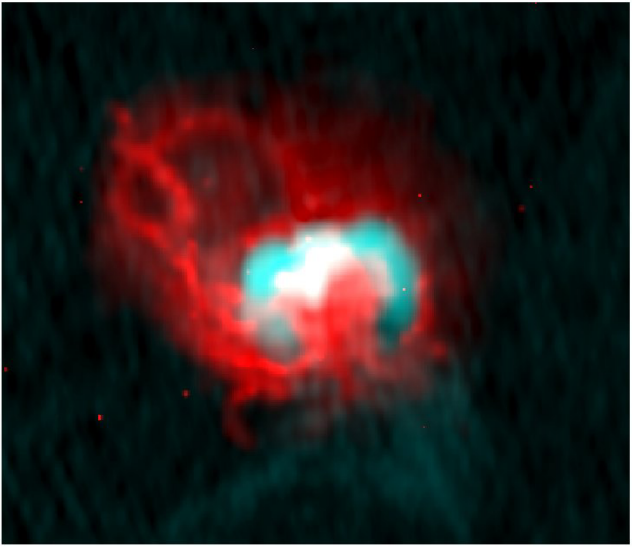}
  \caption{Adaptively-smoothed whole band X-ray image (0.4--7~keV) in
red with overlaid 1.4~GHz radio image in blue. The image is 125 arcsec
from top to bottom.}
\end{figure}

\begin{figure}
  \includegraphics[width=\columnwidth]{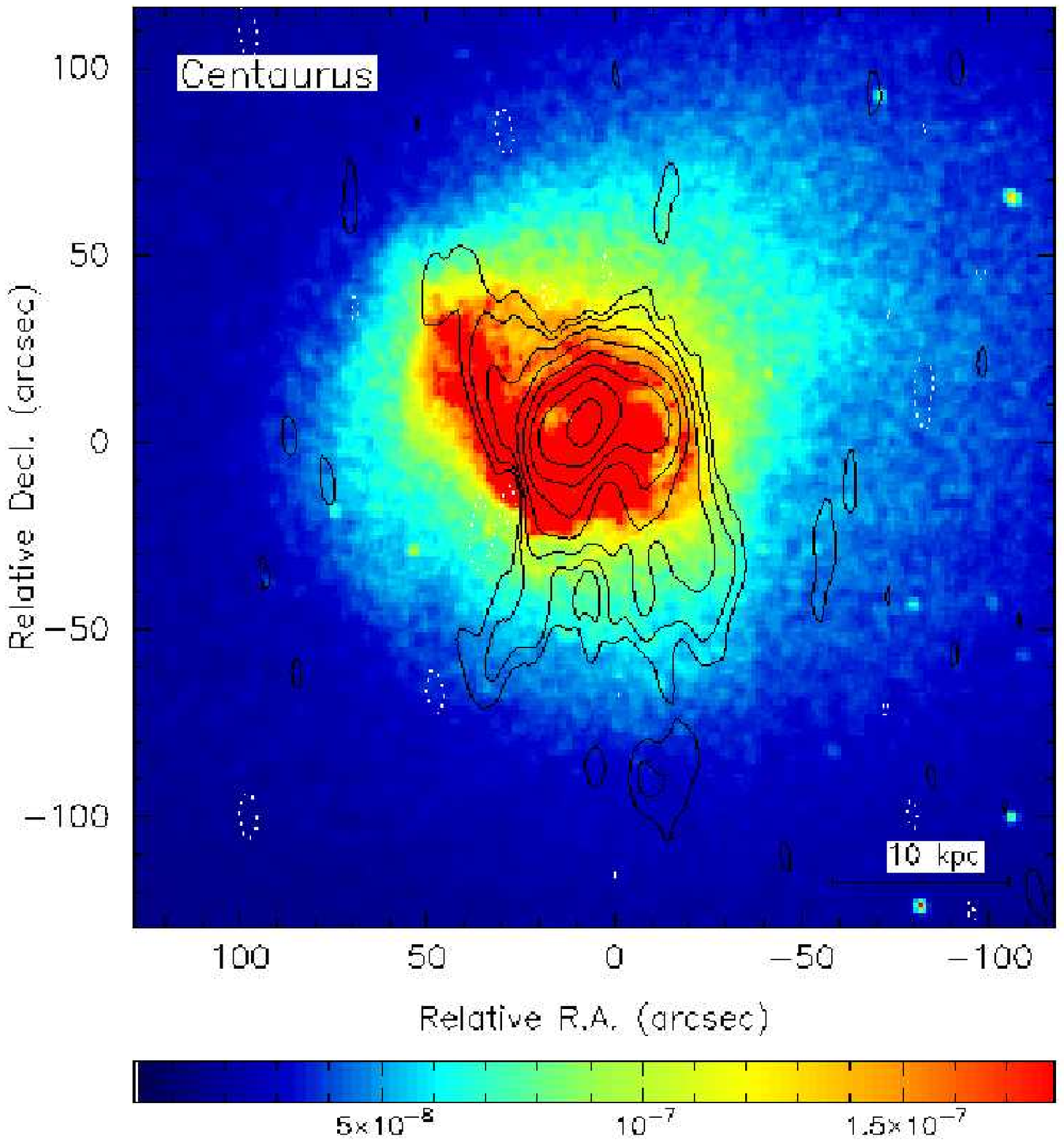}
  \includegraphics[width=\columnwidth]{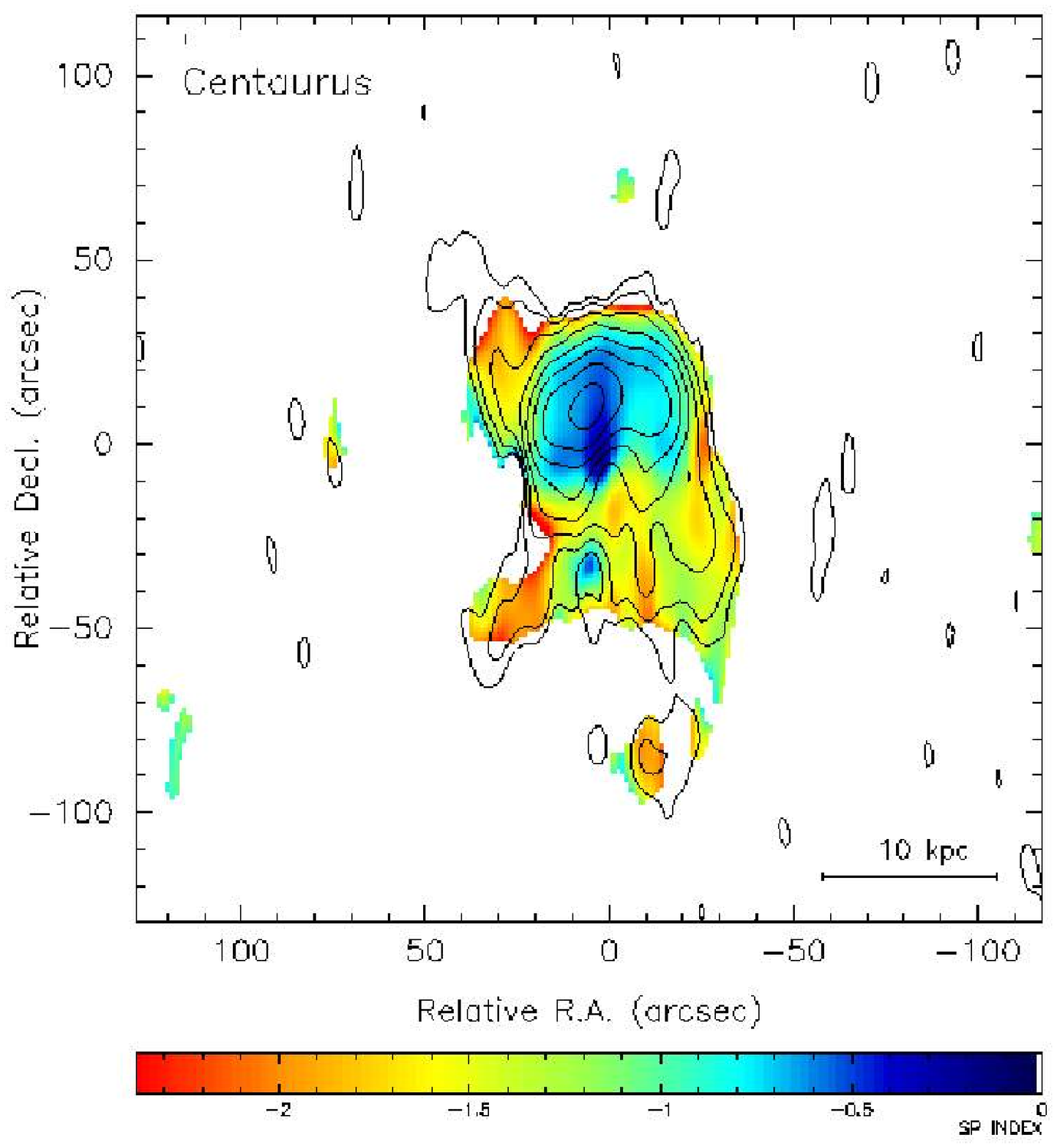}
  \caption{The VLA total intensity 
image at 330 MHz is shown overlaid on the X-ray image (upper) 
and on a spectral index map (lower) with contours 
starting at 20 mJy/beam and increasing by factors of 2.  The 
restoring beam is 17.7 $\times$ 5.6 arcseconds in position angle $-$4.3
degrees.  The radio spectral index (defined $S_\nu \propto 
\nu^\alpha$) has been computed between 330 and 1565 MHz after 
convolving the 1565 MHz image to the resolution of the 330 MHz
image.  }
\end{figure}

$\bullet$ A loop of gas to the SE (Fig.~2, left) does not coincide with
any obvious radio structure (Figs.~7 and 8) and could either surround
relativistic plasma where the electrons have aged to the extent that
they no longer radiate in the observed bands (a ghost bubble), or
could just be a loop of cooler X-ray gas. A slight deficit in this
region seen in the harder X-ray bands (Figs.~1 and 2) suggests that it
is most likely a ghost bubble. A larger loop to the NE appears in the
adaptively-smoothed image (Fig.~7). Radio emission is leaking into the
inner end (Fig.~8), suggesting that it too is a ghost bubble.  
 
The lop-sided larger scale diffuse emission in the total 0.4--7~keV
band was apparent in the earlier image but is now seen much more
clearly. It shows a semicircular edge to the E at 83~arcsec (17.5~kpc)
from the nucleus (Fig.~3) and a larger semicircular edge to the W at
153~arcsec (32.1~kpc). While the edge to the E is centred on the
nucleus of NGC\,4696, that to the W is centred 52~arcsec (10.9~kpc) to
the W of the nucleus, suggesting that the gas there is in motion with
respect to the nucleus. 

The edges are clearly evident in the projected temperature map
(Fig.~4) and emphasise that the temperature abruptly jumps there by
almost 1~keV. They appeared in the radial temperature profiles made
from the earlier data (Sanders \& Fabian 2002) to both E and W
separately. The simplest explanation for them is as cold fronts, such
as seen in other clusters (e.g. Markevitch 2000; Vikhlinin et al
2002). Separate deprojection of spectra across the edges to the E and
W shows that the pressure across the edges is consistent with being
continuous. The abundance drops across both edges; steeply in the case
of the W edge (Fig.~4), unlike the cold front in A496 (Dupke \&
Bregman 2003). The sharp abundance change is similar to behaviour seen
in NGC\,507 by Kraft et al (2004), although the edge there coincides
with a radio lobe, which is not the case here. The large extent of the
edges argues against them being simple cold fronts. Sloshing motions
of the gas within the central potential well (e.g. Markevitch et al
2001) could account for them and for the E-W asymmetry of the
appearance.  However, the semicircular shape of the eastern one (see
particularly Fig. 3) suggests that this, at least,
may have been shaped by disturbances from the nucleus.

The radio source on the other hand appears to be moving to the N,
although that may just be an impression from its outer structure
(Fig.~8) which may be following the path of least resistance (the
optical image of NGC\,4696 shows it significantly flattened in the N-S
direction).

The temperature and abundance maps (Fig.~4) reveal that the core of
the cluster consists of at least 4 distinct parts. They are a) the
region immediately around the nucleus, b) a high abundance region
extending 18~kpc to the end of the eastern plume, c) a quasi-spherical
region of radius 32~kpc with its centre displaced 11~kpc to the W of
the nucleus, and d) the outer region. Each part occupies a different
part of the abundance--temperature plane (Fig.~5) with a) $<1$~keV, b)
$1-2$~keV, c) $2-3$~keV and finally d) $>3$~keV.  

In the present multi-temperature analysis, the abundance peaks at
$\sim2.5\Zsun$ in b) and drops by about $1\Zsun$ between b) and c) and
c) and d). There cannot be any large turbulent or stochastic motions
in the core or these high abundances would have been smoothed out.
Following the work of Rebusco et al (2005), who have estimated the
diffusion coefficient $D$ in the core of the Perseus cluster. we can
obtain a limit from the much higher gradient seen in Centaurus.
Assuming stochastic diffusion only, then steep gradients over a
distance $d$ are smoothed out on a timescale of $d^2/D$. If we first
look at the outer, western edge and rely only on diffusion to smooth
it and adopt a minimum time equal to the local crossing time of the
structure of $\sim 5\times 10^8\yr$ then $D<6\times 10^{28}\cmsqps.$
This is smaller than the result, $D\sim 2\times 10^{29}\cmsqps,$ of
Rebusco et al (2005). Since the region is displaced from the central
galaxy it is unlikely that any current enrichment has a noticeable
effect. Detailed modelling including metal enrichment should yield a
tight constraint from the inner, high abundance, region. The above
constraint from the outer region means that the level of turbulence
and stochastic motions in the Centaurus cluster is low (the product of
velocity and lengthscale of the motions being $\sim 3D$). Smallscale
motions need continual pumping or they would rapidly die out,
temporarily heating the gas, as noted by Rebusco et al (2005).

The lowest energy image reveals delicate filaments of X-ray emitting
gas which are only 1--3~arcsec (200--600~pc) wide. The temperature of
these structures is about 0.7~keV while the gas they are embedded in has
a temperature  of about 1.5~keV. Conduction is presumably much reduced
and magnetic fields help maintain the integrity of the structures.

In the very core around the nucleus, we can now see that the
temperature drops down to less than 0.7~keV, with the inner region
being multiphase. The whole temperature range
detectable in the Centaurus cluster therefore exceeds a factor of 5,
which is significantly larger that the factor of three seen in many
other clusters (Peterson et al 2003; but see Morris \& Fabian 2004).  

A disc galaxy to the SE of NGC\,4696 (at RA 12 49 03.8, Dec -41 20 27,
J2000.0) can just be seen in silhouette in Fig.3 and, more clearly, in
Fig.~6. It has a radial heliocentric velocity of $3737\kmps$
(Dickens, Currie \& Lucey 1986), so lies in the Cen30 cluster, and is
322-G93 in the ESO/Uppsala catalogue. Photoelectric absorption by
galaxies projected onto intracluster gas is also seen in the Perseus
(Gillmon et al 2003) and A2029 (Clarke et al 2004) clusters.

\section{Discussion} 

The inner region is complex with radio bubbles, ghost bubbles and cool
filaments. The 'frothy' X-ray appearance of the centre resembles the
centre of the Virgo cluster (Young et al 2002; Forman et al 2003)
which also has several inner bubbles.  The region beyond is smoother
but appears different to the East and  West. In those directions
clear semicircular edges are seen. The one to the E is concentric with
the nucleus indicating that it may have been triggered by a disturbance
from there. The larger one to the W is centred west of the nucleus.
Its shape suggests that some disturbance from the nucleus may produce
it as well.

The Western edge appears as a marked abundance drop, which is not
easily accounted for, although large buoyant bubbles dragging
iron-rich mattter outward, as seen in the Perseus cluster (Sanders,
Fabian \& Dunn 2004), or fast-moving, outer gas sweeping past as a
result of the ongoing merger with Cen\,45 (which is not otherwise
seen), could contribute. 

The dense, metal-rich centre of the Centaurus cluster is revealed as
more complex than in previous observations. The radio source clearly has an
impact immediately around the centre in the frothy bubble structure
and the Eastern edge, although the long-term nature of the interaction
and its contribution to the energy balance of the central intracluster
medium is elusive. Given that in equilibrium temperature contours
follow equipotentials, the asymmetric temperature structure
around the central galaxy NGC\,4696 means that some major part of the
mass or gas distribution there is dynamically changing.

\section{Acknowledgements} The National Radio Astronomy Observatory is
operated by Associated Universities, Inc., under cooperative agreement
with the National Science Foundation. GBT acknowledges support for
this work from the National Aeronautics and Space Administration
through Chandra Award Number GO4-5135X issued by the Chandra X-ray
Observatory Center, which is operated by the Smithsonian Astrophysical
Observatory for and on behalf of the National Aeronautics Space
Administration under contract NAS8-03060. GBT also thanks the
Institute of Astronomy for hospitality while working on this project.
ACF and SWA thank the Royal Society for support.

\end{document}